\documentclass{iau}
\usepackage{graphicx}

\title[The triple-mode variable AC And] %% give here short title %%
{The rich frequency spectrum of the triple-mode variable AC And}

\author[G. Kov\'acs, G.~A. Bakos \& J.~D. Hartman]   %% give here short author list %%
{G\'eza Kov\'acs$^{1,2}$, G\'asp\'ar A. Bakos$^3$ 
 \and Joel D. Hartman$^3$}

\affiliation{$^1$Dept. of Physics \& Astrophysics, U. North Dakota, Grand Forks, ND\\
[\affilskip]
$^2$Konkoly Obs., Budapest, Hungary \\
[\affilskip]
$^3$Princeton University, Department of Astrophysical Sciences, Princeton
}

\pubyear{2013}
\volume{301}  %% insert here IAU Symposium No.
\pagerange{1--2}
\jname{Precision Asteroseismology}
\editors{J.A. Guzik, W.J. Chaplin, G. Handler \& A. Pigulski, eds.}
\begin{document}

\maketitle

\begin{abstract}
Fourier analysis of the light curve of AC And from the HATNet 
database reveals the rich frequency structure of this object. 
Above $30$ components are found down to the amplitude of 
$3$~mmag. Several of these frequencies are not the linear 
combinations of the three basic components. We detect period 
increase in all three components that may lend support to the 
Pop I classification of this variable.  
\end{abstract}

%\section{Introduction}
There are a handful of objects that seem to exhibit sustained 
pulsations in three radial modes (\cite{wils08}). These are 
important objects with respect to the unique opportunity to derive 
their basic physical parameters (i.e., mass, luminosity and 
temperature) by using their periods only (assuming that the 
metallicity is known and that linear pulsation periods are close 
to the observed [nonlinear] ones -- see \cite[Kovacs \& Buchler (1994)]{kovacs94} and 
\cite{moskalik05}).

%\section{Analysis of the HATNet data}
By using the $9600$ datapoint light curve gathered by 
HATNet\footnote{The Hungarian-made Automated Telescope Network 
(\cite{bakos04}) consists of 6 wide field-of-view, small aperture 
autonomous telescopes located in Hawaii (Mauna Kea) and Arizona 
(Fred Lawrence Whipple Observatory). The prime purpose of the 
project is to search for extrasolar planets via transit technique.}, 
we perform a Fourier frequency analysis on AC~And, the prototype of 
radial three-mode pulsators. The analysis of the basic 
photometric data yields the three modes and their low-order linear 
combinations with high S/N. To reach the millimagnitude level, we 
need to minimize systematics. Because of the high amplitudes of 
the low-order components, we cannot use TFA (\cite{kovacs05}) in the 
frequency search mode, since it is based on the low S/N assumption. 
Therefore, we opted to use the method of \cite{kovacs08}. Here the most 
reliably determined {\em signal components and the TFA template 
light curves are simultaneously fitted}. After cleaning the data 
by the so-obtained TFA filter, we get clean, successively 
prewhitened spectra as shown on the left panel of Fig.~1. Successive 
prewhitening leads to the identification of many components, most 
of which are clearly the linear combinations of the 3 basic 
frequencies. However, both the frequency spectra (top right panel 
of Fig.~1) and the decomposed individual modes show the presence 
of non-fitting components even when considering linear combinations 
of the main frequencies up to order $10$. This feature has also been 
noted in the discovery paper of \cite{fitch76}. 

%
%%%%%% FIG. 1
%
\begin{figure}[t]
\vspace*{0mm}
\begin{flushleft}              
\includegraphics[angle=0,width=70mm]{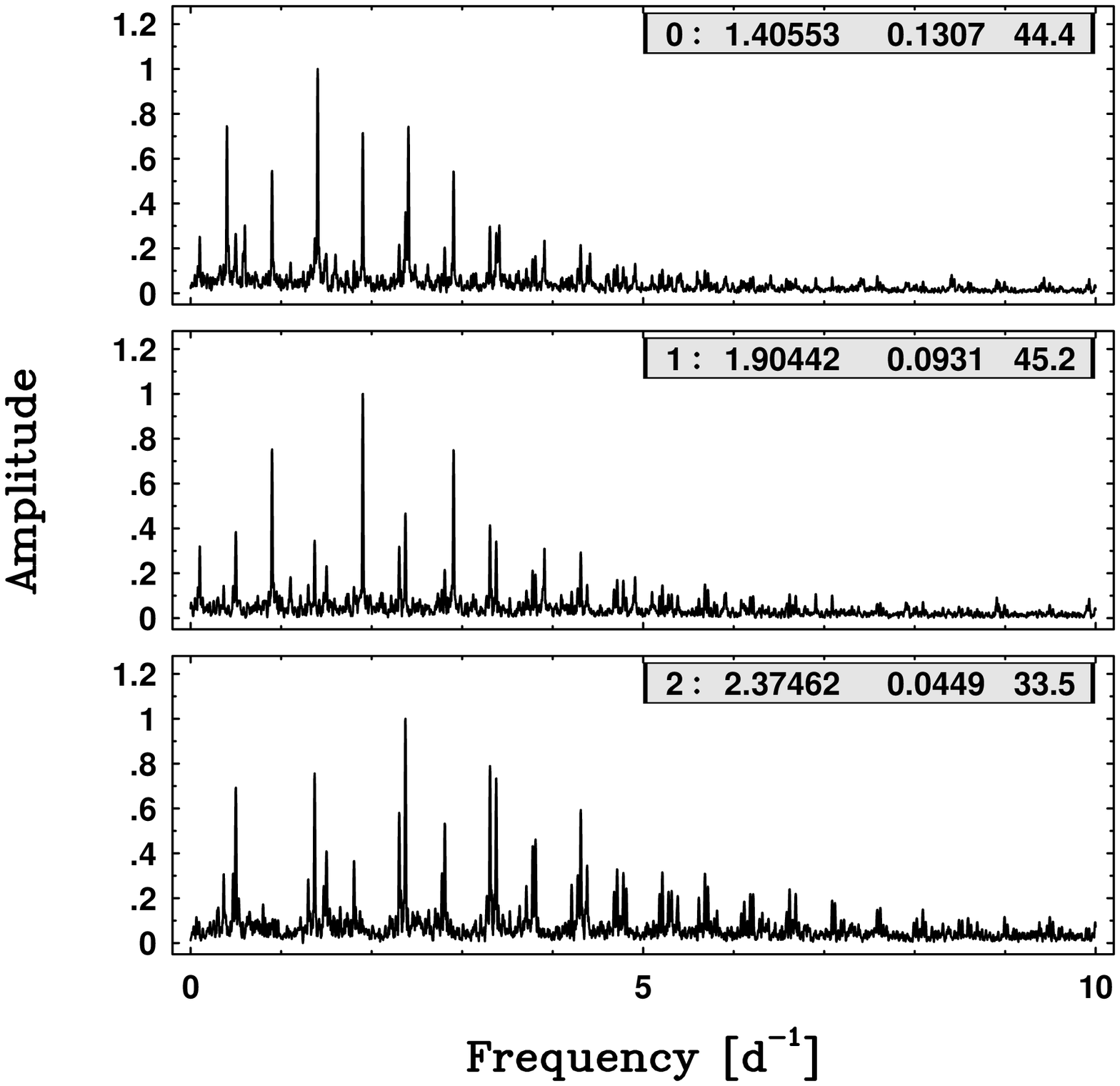}
\end{flushleft}
\vspace*{-68.5mm}
\begin{flushright}              
\includegraphics[angle=0,width=65mm]{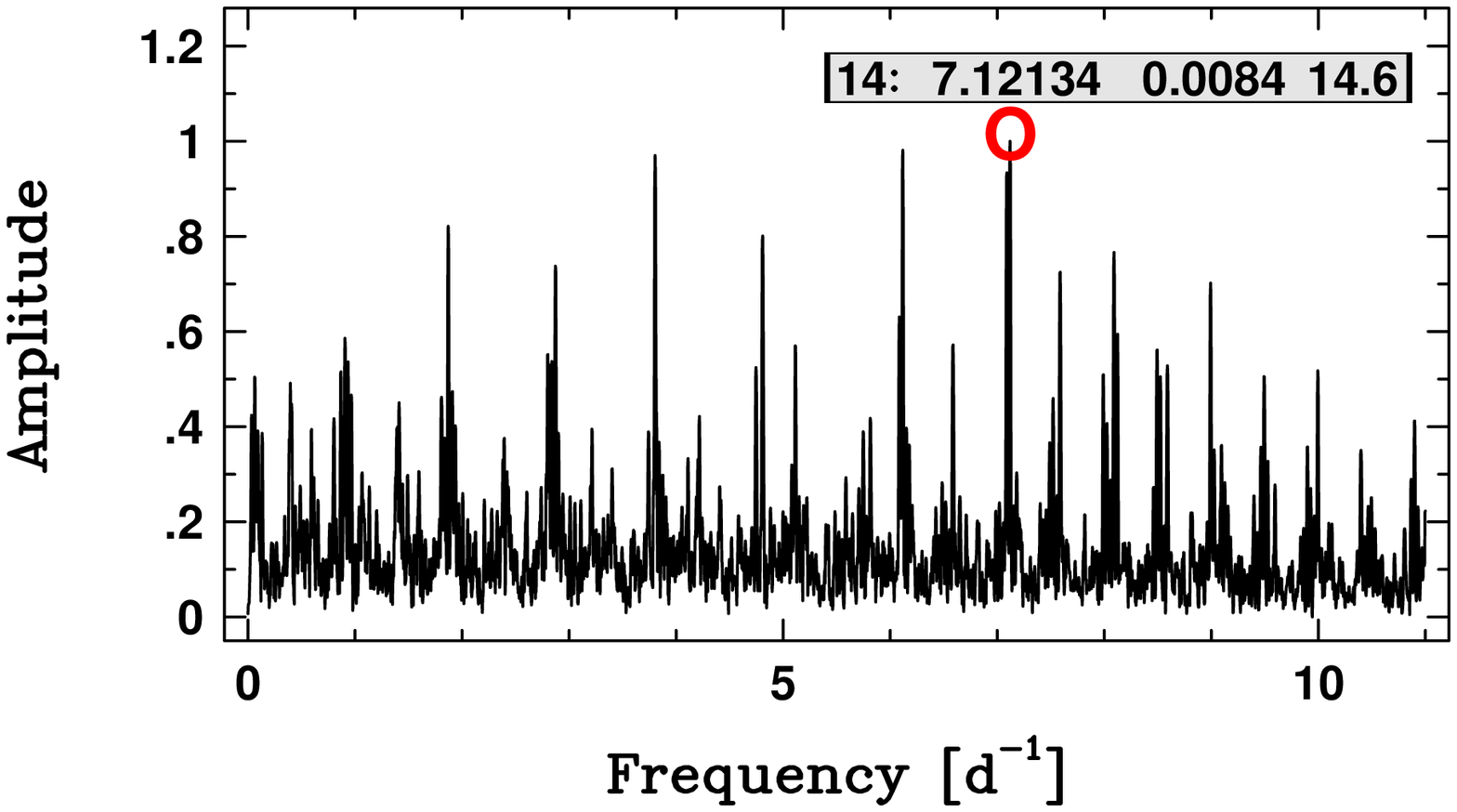}
\end{flushright}
\vspace*{-0mm}
\begin{flushright}              
\includegraphics[angle=0,width=65mm]{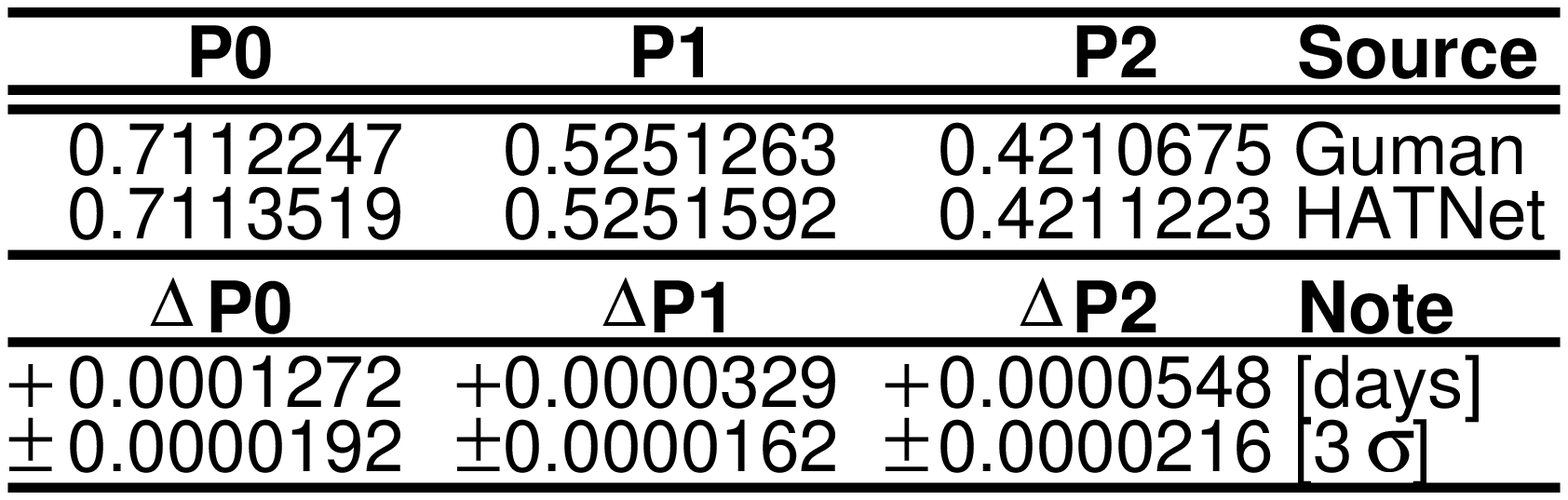}
\end{flushright}
\vspace*{7mm}
\caption{{\it Left:} Successively prewhitened frequency spectra of AC~And; 
               insets: prewhitening order; peak frequency [c/d], 
	       amplitude [mag], S/N.
         {\it Top right:} Frequency spectrum of AC~And after 
	        the 14th prewhitening. The circled peak corresponds 
		to one of the handful of components that cannot be 
		identified as the linear combination of the three main 
		frequencies.  
         {\it Bottom right:} Periods detected 55~years apart 
	 (upper two rows) and their differences together with the 
	 corresponding errors (lower two rows). 
}
\end{figure}
%

%\section{Period changes, physical parameters}
We also examined the rate of period change during the past 55~years. 
Knowledge of the period change is important, because if it is due to 
stellar evolution, it may help to select the right model if other 
parameters are ambiguous. From the analysis of \cite[Guman (1981, 1982)]{guman81,guman82} it 
seems that all three modes exhibit period increases much higher than 
expected from an RR~Lyrae star (\cite{kovacs94}). As seen in the 
table section of Fig.~1, the periods derived from the HATNet data 
confirm the period increase during the past 55 years. Note however 
that the speed of increase derived here is $\sim4$ times higher 
than given by \cite{jurcsik06} based on earlier data.  

Except perhaps for the Str\"omgren photometry of \cite{pena05}, AC~And 
lacks good multicolor time series observations and deep spectroscopic 
work. As a result, the physical parameters of the star are still 
poorly known. Opposite to \cite{pena05}, we think that the 
available photometric and pulsation data support that AC~And is a 
relatively high ($\sim 3$~M$_{\odot}$) mass star with 
$T_{\rm eff}\approx 5800\pm200$~K, $\log g\approx 2.5\pm0.5$ and 
[Fe/H] $\approx -0.5\pm0.5$. 

\vskip 3mm
\noindent
{\em Acknowledgement:} 
G.~K. thanks the Hungarian Scientific Research Foundation (OTKA) 
for support through grant K-81373.

\end{document}